\documentclass[11pt]{article}
\usepackage[margin=1in]{geometry}
\usepackage{amsmath}
\usepackage{amssymb}
\usepackage{amsthm}
\usepackage{bm}
\usepackage{graphicx}
\usepackage{subfig}
\usepackage{cite}
\usepackage[usenames,dvipsnames]{xcolor}
\usepackage[hidelinks]{hyperref}
\usepackage[capitalise]{cleveref}

\title{A Finite Population Destroys a Traveling Wave in Spatial Replicator Dynamics}

\author{Christopher Griffin
\footnote{Applied Research Laboratory, Penn State University,
University Park, PA 16802}~\footnotemark[3]
\and
Riley Mummah\footnote{Department of Ecology and Evolutionary Biology,
UCLA, Los Angeles, CA 90095}
\and
Russ deForest
\footnote{Department of Mathematics, Penn State University, University Park, PA 16802}
}

\date{}

\begin{document}

\maketitle

\begin{abstract}
We derive both the finite and infinite population spatial replicator dynamics as the fluid limit of a stochastic cellular automaton. The infinite population spatial replicator is identical to the model used by Vickers and our derivation justifies the addition of a diffusion to the replicator. The finite population form generalizes the results by Durett and Levin on finite spatial replicator games. We study the differences in the two equations as they pertain to a one-dimensional rock-paper-scissors game. 
\end{abstract}


\section{Introduction}
Evolutionary games using the replicator dynamic have been studied extensively and are now well documented \cite{ref:Fr91,ref:Wei95,ref:HS98,ref:HS03,T15,T19}. Variations on the classical replicator dynamic include discrete time dynamics \cite{ref:VRS11} and mutations \cite{ref:BTG12,ref:TS15}. Additional evolutionary dynamics, such as imitation  \cite{ref:Fr91,ref:Sc97,ref:FL98,ref:HS03} and exchange models \cite{ref:EGB16} have been studied. Alternatively evolutionary games have been extended to include spatial models by a number of authors \cite{ref:andrew,ref:kerr,ref:nowak,ref:roca,ref:RPSnetworks,ref:vickers1989,ref:vickers1991,ref:cressmanVickers}. Most of these papers append a spatial component to the classical replicator dynamics (see e.g., \cite{ref:vickers1989}) or discuss finite population replicator dynamics in which total population counts are used (see e.g., \cite{ref:andrew}). In the latter case, a spatial term is again appended to the classical replicator structure.

In \cite{DL94}, Durrett and Levin study discrete and spatial evolutionary game models and compare them to their continuous, aspatial analogs. For their study the authors focus on a specific class of two-player two-strategy games using a hawk-dove payoff matrix. Because their payoff matrix is $2 \times 2$, a single (spatial) variable $p(x,t)$ can be used to denote the proportion of the population playing hawk (Strategy 1) while a second variable $s(x,t)$ denotes the total population. Using the payoff matrix:
\begin{displaymath}
  \mathbf{A} = \begin{bmatrix}a & b\\c&d\end{bmatrix},
\end{displaymath}
the remarkable reaction-diffusion equation is analyzed: 
\begin{align}
  \frac{\partial p}{\partial t} &= \Delta p + \frac{2}{s}\nabla p\cdot \nabla s + pq((a-c)p + (b-d)(1-p))
  \label{eqn:Durret1}
  \\
  \frac{\partial s}{\partial t} &= \Delta s + s\left(ap^2 + (b+c)p(1-p)+(1-p)^2d\right) - \kappa s^2.
  \label{eqn:Durret2}
\end{align}
Here $\kappa$ is a death rate due to overcrowding and $q = (1-p)$. Let $\mathbf{e}_i \in \mathbb{R}^n$ be the $i^\text{th}$ unit vector. Let $\mathbf{u} = \langle{p,q}\rangle$. When $\kappa = 0$, we can rewrite these equation as:
\begin{align}
  \frac{\partial p}{\partial t} &= \Delta p + \frac{2}{s}\nabla p\cdot \nabla s + p\left(\mathbf{e}_1^T-\mathbf{u}^T\right)\mathbf{A}\mathbf{u}\label{eqn:DL1}\\
  \frac{\partial s}{\partial t} &= \Delta s + s\cdot\mathbf{u}^T\mathbf{A}\mathbf{u}.
\end{align}
That is, Durrett and Levin have encoded the replicator dynamic into a finite population spatial partial differential equation, which differs from the one used by Vickers \cite{ref:vickers1989} because it assumes a finite population in its derivation. The second term in \cref{eqn:DL1} is an advection of species $p$ along a velocity vector given by the gradient of the whole population. However, this advection speed is inversely proportional to the population size. Thus small populations can have a dramatic effect on this term. As we show in \cref{sec:FluidLim}, this behavior  carries through for general games. To illustrate the advection clearly, we can define the potential function:
\begin{displaymath}
\phi(x) = 2\log\left[s(x,t)\right].
\end{displaymath}
We then have:
\begin{displaymath}
\mathbf{v} = \nabla\phi = \frac{2}{s}\nabla s.
\end{displaymath}
\cref{eqn:DL1} can then be written as:
\begin{displaymath}
\frac{\partial p}{\partial t} = \Delta p + \mathbf{v}\cdot \nabla p + p\left(\mathbf{e}_1^T-\mathbf{u}^T\right)\mathbf{A}\mathbf{u}\label{eqn:DL1a},
\end{displaymath}
giving a standard advection-diffusion equation. The logarithm in the potential function conveniently yields the per-capita advection rate.

In this paper we show that Durrett and Levin's finite population
spatial replicator is generalizable to an arbitrary payoff matrix. We
then focus our attention on the one-dimensional rock-paper-scissors
game under the replicator dynamics in both finite and infinite populations. This system has interesting properties in both the finite and infinite
population cases. In particular, we show: (i) the  model used by
Vicker's \cite{ref:vickers1989} arises naturally as the infinite
population limit of the generalization of Durrett and Levin's model,
which in turn can be derived from a stochastic cellular automaton
(particle) model as a fluid limit, and (ii) the one dimensional infinite population spatial rock-paper-scissors dynamic has a constant amplitude traveling wave solutions in biased RPS. However, the finite population version does not exhibit such solutions, but does seem to exhibit attracting stationary solutions. We illustrate the latter result numerically.

\section{Model}\label{sec:Model}
Let $\mathbf{A} \in \mathbb{R}^{n \times n}$ be a payoff matrix for a symmetric game \cite{ref:HS03}. All vectors are column vectors unless otherwise noted. Below we construct a stochastic cellular automaton and show that the fluid limit of this system yields a generalization of Durrett and Levin's specific finite population spatial replicator.

The state of cell $i$ at time index $k$ of the cellular automaton is a tuple $\langle{U_1(i,k),\dots,U_n(i,k)}\rangle$ where $U_j(i,k)$ provides the size of the population of species $j$ at position $i$ at time $k$. For simplicity, we assume that species interaction may only happen between cells and not within cells; i.e., the $U_1(i,k)$ members of species $1$ will not play  against the $U_n(i,k)$ members of species $n$. This assumption will become irrelevant in the limit.

\begin{figure}[htbp]
\centering
\includegraphics[width=0.4\columnwidth]{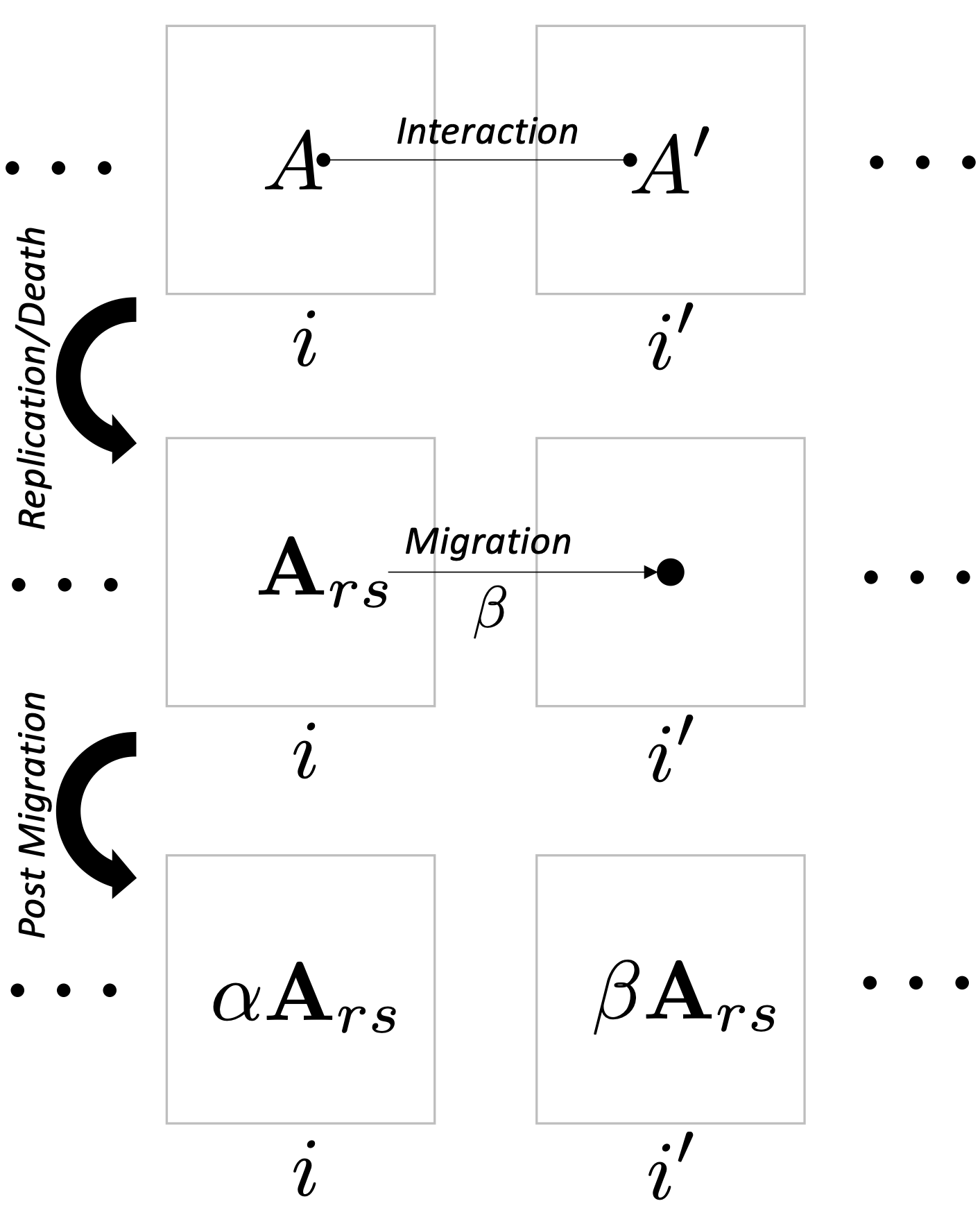}
\caption{A single interaction between two agent in a spatial game on
  a one-dimensional lattice is illustrated. The direction of interaction is chosen at random. Motion is random and governed by $\alpha$ and $\beta$ where $\alpha+\beta=1$.}
\label{fig:Update}
\end{figure}

During state update, an agent $A$ at cell $i$ chooses a random direction (cell $i'$ ) and a random member of the population (agent $A'$) within that cell. Assume Agent $A$ uses strategy $r$ while Agent $A'$ uses strategy $s$. After play, there are $\alpha\cdot\mathbf{A}_{rs}$ additional agents at cell $i$ using strategy $r$ and $\beta\cdot\mathbf{A}_{rs}$ additional agents playing strategy $r$ at cell $i'$, where $\alpha + \beta = 1$ is the probability of motion from cell $i$ to cell $i'$.
If $\mathbf{A}_{rs} < 0$, then agents are removed from their respective cells. To avoid computational issues when more members of a species die than are present, $\mathbf{A}$ can be modified so that $\mathbf{A}_{rs} \geq 0$ for all $r,s \in \{1,\dots,n\}$, without altering the evolutionary dynamics in proportion \cite{ref:Wei95}. The update rule for a single agent is illustrated in \cref{fig:Update}.
The process described above and illustrated in \cref{fig:Update} is assumed to be happening simultaneously for each agent and we assume that the replication/death as a result of game play along with the migration are happening on (roughly) the same time scale. 
%
%
Using these assumptions, we can construct mean-field equations for the population $U_r$ at position $i$ and time $k+1$ in a 1-D cellular automaton, assuming an equal likelihood that agents diffuse left or right. The mean number of agents playing strategy $r$ present at position $i$ at time $k+1$ is determined by:
\begin{enumerate}
\item The expected number of agents who remain at cell $i$: $\alpha U_r(i,k)$
\item The expected number of new agents created at cell $i$ who remain at cell $i$:
\begin{displaymath}
\frac{\alpha}{2}U_r(i,k)\left(\sum_{s}\mathbf{A}_{rs}\left(u_s(i+1,k) + u_s(i-1,k)\right)\right).
\end{displaymath}
\item The expected number of agents who migrate to position $i$ from neighboring cells:
\begin{displaymath}
\frac{\beta}{2}\left(U_r(i+1,k)+U_r(i-1,k)\right).
\end{displaymath}
\item The expected number of agents created in a neighboring cell who migrate to cell $i$:
\begin{displaymath}
\frac{\beta}{2}\left(U_r(i+1,k)+U_r(i-1,k)\right)\sum_{s}\mathbf{A}_{rs}u_s(i,k).
\end{displaymath}
\end{enumerate}
Here $u_s(i,k)$ is the \textit{proportion of the population} at cell $i$ playing strategy $s$ at time step $k$.

Let $\mathbf{u}(i,k) = \langle{u_1(i,k),\dots,u_n(i,k)}\rangle$. Re-writing sums as matrix products, the expected number of agents playing strategy $r$ at cell $i$ at time $k+1$ is:
\begin{multline}
U_r(i,k+1) =
\alpha U_r(i,k) + \frac{\beta}{2}\left(U_r(i+1,k)+U_r(i-1,k)\right) +\\
\frac{\alpha}{2}U_r(i,k)\left(\mathbf{e}_r^T\mathbf{A}\left(\mathbf{u}(i+1,k) + \mathbf{u}(i-1,k)\right)\right) + \\
\frac{\beta}{2}\left(U_r(i+1,k)+U_r(i-1,k)\right)\mathbf{e}_r^T\mathbf{A}\mathbf{u}(i,k).
\label{eqn:MatrEqn}
\end{multline}

Assume the cellular grid has lattice spacing $\Delta x$. Following \cite{ref:DGB+14} and using a Taylor approximation, we can write:
\begin{align*}
    U_r(x,t+\Delta t) &\approx U_r(x,t) +
        \Delta t \frac{\partial U_r(x,t)}{\partial t} + O(\Delta t^2)\\
    U_r(x+\Delta x,t) &\approx \sum_{j=0}^{2}\frac{\Delta x^j}{j!}
        \frac{\partial^j U_r(x,t)}{\partial x^j} + O(\Delta x^3).
\end{align*}

\section{Derivation of Fluid Limits}\label{sec:FluidLim}
We proceed to derive the mean-field approximation. Passing to the continuous case and assuming that interaction rates decrease linearly with $\Delta t$, we can write a second order approximation of \cref{eqn:MatrEqn} as:
\begin{multline}
\Delta t\frac{\partial U_r(x,t)}{\partial t} = \alpha U_r(x,t) + 
\beta\left(U_r(x,t) + \frac{1}{2}\Delta x^2\frac{\partial^2 U_r}{\partial x^2}\right) + \\
\alpha\Delta t U_r(x,t)\cdot \mathbf{e}_r^T\mathbf{A}\left(\mathbf{u}(x,t) + \frac{1}{2}\Delta x^2\frac{\partial^2 \mathbf{u}}{\partial x^2} \right) + \\
\beta\Delta t\left(U_r(x,t) + \frac{1}{2}\Delta x^2\frac{\partial^2 U_r}{\partial x^2}\right)\mathbf{e}_r^T\mathbf{A}\mathbf{u}(x,t) - U_r(x,t).
\end{multline}
Where the $-U_r(x,t)$ on the right-hand-side arises from the formation of the Newton quotient on the left-hand-side. 
Expanding and simplifying  yields:
\begin{multline}
\Delta t\frac{\partial U_r(x,t)}{\partial t} =
\frac{\beta}{2}\Delta x^2\frac{\partial^2 U_r}{\partial x^2} +
\Delta t U_r(x,t)\mathbf{e}_r^T\mathbf{A}\mathbf{u}(x,t) + \\
\frac{\alpha\Delta t}{2} \Delta x^2 U_r(x,t)\mathbf{e}_r^T\mathbf{A}\frac{\partial^2 \mathbf{u}}{\partial x^2} + \\
\frac{\beta\Delta t}{2}\Delta x^2\frac{\partial^2 U_r}{\partial x^2}\mathbf{e}_r^T\mathbf{A}\mathbf{u}(x,t).
\end{multline}
Assume $\beta \in (0,1]$. Dividing through by $\Delta t$ and assuming that $\lim_{\Delta t \rightarrow 0}\Delta x^2/\Delta t = 2D/\beta$ yields:
\begin{equation}
\frac{\partial U_r(x,t)}{\partial t} =  U_r(x,t)\mathbf{e}_r^T\mathbf{A}\mathbf{u}(x,t) +
D\frac{\partial^2 U_r}{\partial x^2}.
\label{eqn:righteqn}
\end{equation}
The constant $D$ is the diffusion constant and the assumption that 
$$\lim_{\Delta t \rightarrow 0}\Delta x^2/\Delta t = 2D/\beta$$
is a variant of the assumption used to derive Fick's Law \cite{O09} and identical when $\beta = 1$.

These are the spatial dynamics used by Durrett and Levin, (in the first part of their paper), but are derived only by adding a diffusion term to the standard finite population growth equations. In \cite{DL94}, Durrett and Levin note that they derive a set of equations they feel are more appropriate for modeling finite spatial systems. Their derivation at the end of \cite{DL94} (for a specific hawk-dove system) rests on the assumption that migration happens ``on a much faster timescale'' than game interactions. Our model assumes that migration and game interactions occur on approximately the same time scale. Under this assumption, \cref{eqn:righteqn} is the correct spatial adaptation for finite populations; i.e., one simply adds a diffusion term. In the case where migration happens more quickly, then the derivation in \cite{DL94} should be used instead.

\section{Spatial Replicator with Finite Population}
The derivation of \cref{eqn:Durret1,eqn:Durret2} are not given in \cite{DL94}. They can be generalized for an arbitrary evolutionary game using \cref{eqn:righteqn} as the starting point. Let:
\begin{displaymath}
M(x,t) = \sum_{s}U_s(x,t).
\end{displaymath}
Differentiating we have:
\begin{displaymath}
\frac{\partial}{\partial t}\frac{U_r(x,t)}{M(x,t)} = \frac{1}{M(x,t)}\frac{\partial U_r(x,t)}{\partial t} - \\
u_r(x,t)\sum_{s}\frac{1}{M(x,t)}\frac{\partial U_s(x,t)}{\partial t}.
\end{displaymath}
Substituting from \cref{eqn:righteqn} we obtain:
\begin{equation}
\frac{\partial u_r}{\partial t} = u_r\cdot\left(\mathbf{e}_r^T\mathbf{A}\mathbf{u} -\mathbf{u}^T\mathbf{A}\mathbf{u}\right)+\\
\frac{D}{M}\left(\frac{\partial^2 U_r}{\partial x^2} - u_r\cdot\sum_{s}\frac{\partial^2 U_s}{\partial x^2} \right).
\label{eqn:PartialResult}
\end{equation}
Unlike in the derivation of the standard replicator dynamic, the rate of change of the population proportion is not solely a function of the proportions themselves. 

We can remove dependence on the individual populations to derive an independent (coupled) system of differential equations that includes only the total population. 
For arbitrary strategy $r$, we can apply the quotient rule to obtain:
\begin{displaymath}
M\frac{\partial u_r}{\partial x} = \frac{\partial U_r}{\partial x} - u_r\frac{\partial M}{\partial x}
\end{displaymath}
Differentiating again, multiplying by $1/M$ and re-arranging yields the expression:
\begin{displaymath}
\frac{1}{M}\frac{\partial^2 U_r}{\partial x^2} = \frac{u_r}{M}\frac{\partial^2 M}{\partial x^2} + \frac{2}{M}\frac{\partial u_r}{\partial x}\frac{\partial M}{\partial x} + \frac{\partial^2 u_r}{\partial x^2}.
\end{displaymath}
Using this we can write
\begin{displaymath}
\frac{D}{M}\left(\frac{\partial^2 U_r}{\partial x^2} - u_r\sum_{s}\frac{\partial^2 U_s}{\partial x^2} \right) = 
D\left(\frac{2}{M}\frac{\partial M}{\partial x}\frac{\partial u_r}{\partial x}  + \frac{\partial^2 u_r}{\partial x^2} \right)
\end{displaymath}
Using this we can re-write \cref{eqn:PartialResult} as: 
\begin{equation}
\frac{\partial u_r}{\partial t} = u_r\cdot\left(\mathbf{e}_r^T\mathbf{A}\mathbf{u} -\mathbf{u}^T\mathbf{A}\mathbf{u}\right)+\\
 D\left(\frac{2}{M}\frac{\partial M}{\partial x}\frac{\partial u_r}{\partial x}  + \frac{\partial^2 u_r}{\partial x^2} \right).
\end{equation}
The dynamics of $M$ can be derived (by addition) from \cref{eqn:righteqn}:
\begin{displaymath}
\frac{\partial M}{\partial t} = M\mathbf{u}^T\mathbf{A}\mathbf{u} + D\frac{\partial^2 M}{\partial x^2}.
\end{displaymath}
Thus, we have a coupled set of differential equations written entirely in terms of $\mathbf{u}$ and $M$, rather than $U_r$, $M$ and $\mathbf{u}$:
\begin{equation}
\left\{
\begin{aligned}
&\forall r \left\{\begin{aligned}
\frac{\partial u_r}{\partial t} = & u_r\cdot\left(\mathbf{e}_r^T\mathbf{A}\mathbf{u} -\mathbf{u}^T\mathbf{A}\mathbf{u}\right)+\\
& \hspace*{2em}D\left(\frac{2}{M}\frac{\partial M}{\partial x}\frac{\partial u_r}{\partial x}+ \frac{\partial^2 u_r}{\partial x^2} \right)
\end{aligned}\right.\\
&\hspace*{2em}\frac{\partial M}{\partial t} = M\mathbf{u}^T\mathbf{A}\mathbf{u} + D\frac{\partial^2 M}{\partial x^2}
\end{aligned}
\right..
\label{eqn:finitespatialrep}
\end{equation}
This is the spatial replicator equation for \textit{finite}
populations. Letting $M = s$ and $D = 1$, we recover the dynamics of
Durrett and Levin. In contrast to the aspatial replicator, the
inclusion of dynamics for $M$ yields a linearly independent system of
differential equations.

Allowing $M$ to approach infinity uniformly in $x$, we arrive at the
fluid limit in terms of $\mathbf{u}$ alone; this is the 1D nonlinear reaction-diffusion equation used by Vicker's \cite{ref:vickers1989,ref:vickers1991}:

\begin{equation}
\forall r \left\{
\begin{aligned}
\frac{\partial u_r}{\partial t} &= u_r\cdot\left(\mathbf{e}_r^T\mathbf{A}\mathbf{u} -\mathbf{u}^T\mathbf{A}\mathbf{u}\right) + D\frac{\partial^2 u_r}{\partial x^2}
\end{aligned}
\right..
\label{eqn:spatialrep}
\end{equation}
Generalization to $N$-dimensions is straightforward by replacing $\partial^2_x$ with the Laplacian $\Delta$. The $N$-dimensional spatial replicator with finite population is given by:
\begin{align*}
&\forall r \left\{\begin{aligned}
\frac{\partial u_r}{\partial t} = & u_r\cdot\left(\mathbf{e}_r^T\mathbf{A}\mathbf{u} -\mathbf{u}^T\mathbf{A}\mathbf{u}\right)+ 
\left(\frac{2}{M}\nabla M \cdot \nabla u_r +  \Delta u_r\right)
\end{aligned}\right.\\
&\hspace*{2em}\frac{\partial M}{\partial t} = M\mathbf{u}^T\mathbf{A}\mathbf{u} + D\Delta M.
\end{align*}
Thus, the finite population case adds a nonlinear advection term that forces $u_r$ to follow the population gradient. A similar system is studied by deForest and Belmonte in \cite{ref:andrew}, where the payoff gradient is followed instead of the population gradient.

In both \cref{eqn:finitespatialrep,eqn:spatialrep}, we see that the
aspatial replicator dynamics appear on the right hand side perturbed
by a spatial term. It is well known that the dynamics of the aspatial
replicator are confined to the $n$-dimensional simplex $\Delta_n$.
This remains true for the spatial replicator dynamics with finite populations.
Moreover, the solution $u_r = 1$ (i.e., there is only one population)
is a fixed point for the spatial replicator dynamic since the spatial
derivative of the probability distribution of the population
proportions is zero and the time derivative is identically zero as
expected. Thus, pure populations are constant stationary solutions for
these dynamics. Lastly, if $\tilde{\mathbf{u}}(x,t)$ is a constant
solution at a Nash equilibrium for the game defined by $\mathbf{A}$,
then the right-hand-side is again identically zero by the Folk Theorem
\cite{ref:HS03} of evolutionary game theory together with the fact
that there is no spatial variation. Thus every Nash equilibrium of the
matrix game corresponds to a spatially constant stationary solution of the spatial replicator dynamic in both the finite and infinite population cases. 

\section{Example: One Dimensional Rock-Paper-Scissors}\label{sec:RPS}
Durrett and Levin's analysis of Hawk-Dove was aided by the fact that
one strategy can be eliminated, leaving a coupled system of two
partial differential equations. In the remainder of this paper, we analyze variations of rock-paper-scissors (RPS), which yield more interesting results because of its cyclic three-strategy nature and because it can be easily parameterized as discussed in \cite{ref:Wei95}. Substantial work has been done on (spatial) RPS without assuming a replicator dynamic or assuming a general replicator dynamic \cite{ML75,M10,PR19,SMR14,SMJS14,RMF07,RMF08,HMT10,PR17,SMR13}. In an example of a very recent generalization,  Kabir \& Tanimoto study pairwise evolution in RPS with noise \cite{KT21}.

A major focus of work in non-replicator spatial RPS has been on identifying spatial structures (e.g., spirals) that can emerge from these dynamics. For completeness, we quantify the relationship between the work in \cite{PR19,SMR14,SMJS14,RMF07,RMF08,HMT10,PR17,SMR13} and the replicator dynamic considered in this work in Appendix A. There, we show that the equation system used in these references does not subsume results on the replicator dynamic in general and hence one cannot extend these results automatically to this case.

The objective in the remainder of this work is to compare the finite and infinite population replicator for a more interesting three strategy game. For simplicity (and in contrast to much of the previous work in spatial RPS), we focus our attention on the one-dimensional PDE. 

We consider a generalized RPS payoff matrix as given by:
\begin{displaymath}
\mathbf{A} = \begin{bmatrix}
0 & -1 & 1+a\\
1 +a & 0 & -1\\
-1 & 1+a & 0
\end{bmatrix}.
\end{displaymath}
When $a = 0$, this is the standard RPS game which has Nash equilibrium
$\langle{\tfrac{1}{3},\tfrac{1}{3},\tfrac{1}{3}}\rangle$. This is the
unique interior fixed point and the aspatial replicator exhibits an elliptic fixed point at this Nash equilibrium. This Nash equilibrium is preserved for $a \neq 0$;  and
corresponds to an asymptotically stable interior fixed point when $a
> 0$ and unstable fixed point when $a < 0$ \cite{Ze80} as a result of a degenerate Hopf bifurcation. This degenerate Hopf bifurcation is well understood and leads to traveling wave solutions, which we discuss in \cref{sec:TravelingWave}. We note $\mathbf{A}$ is not as fully general as the RPS matrix given in \cite{ref:HS03} or \cite{Ze80}, but it is substantially easier to work with.

Let $\mathbf{u} = \langle{u_r,u_p,u_s}\rangle$ and note:
\begin{displaymath}
\zeta(u_r,u_p,u_s) \overset{\Delta}{=} \mathbf{u}^T\mathbf{A}\mathbf{u} = a\left(u_ru_s + u_ru_p + u_s u_p\right).
\end{displaymath}
There are at least two classes of global solutions to \cref{eqn:finitespatialrep,eqn:spatialrep}: 
\begin{description}
\item[Equilibrium Solution] Here, $u_r = u_p = u_s = \tfrac{1}{3}$ and $M$ solves:
\begin{displaymath}
0 = \frac{a}{3}M + M''
\end{displaymath}
or there is a single population (e.g., $u_r = 1$) and $M$ satisfies $M''=0$.

\item[Oscillating Solution] Here, $u_*(x,t) \equiv \upsilon_*(t)$ with $* \in \{r,p,s\}$ where $\upsilon_r$, $\upsilon_p$, $\upsilon_s$ are solutions to the standard RPS replicator and $M(x,t) = \mu(t)$ satisfies linear equation:
\begin{displaymath}
\dot{\mu} = \zeta(\upsilon_r,\upsilon_p,\upsilon_s)\mu.
\end{displaymath}
\end{description}
In either case, we may impose appropriate boundary conditions.

\subsection{Traveling Wave Solutions}\label{sec:TravelingWave}
Consider the infinite population model. Letting $z = x - ct$, we can re-write \cref{eqn:spatialrep} in compact form as:
\begin{equation}
\forall i\left\{
\begin{aligned}
Dv_i' &= -u_i\left(\mathbf{e}_i^T\mathbf{A}\mathbf{u} - \mathbf{u}^T\mathbf{A}\mathbf{u}\right) -c v_i\\
u_i' &= v_i
\end{aligned}\right..
\label{eqn:TravelingWave}
\end{equation}
For RPS we have the six dimensional non-linear system:
\begin{equation}
\left\{
\begin{aligned}
Dv_r' &= a u_r\zeta(u_r,u_s,u_p) - u_r(u_s - u_p) - a u_r u_s - c v_r\\
u_r' &= v_r\\
Dv_p' &= a u_p\zeta(u_r,u_s,u_p) - u_p(u_r - u_s) - a u_p u_r - c v_p\\
u_p' &= v_p\\
Dv_s' &= a u_s\zeta(u_r,u_s,u_p) - u_s(u_p - u_r) - a u_s u_p - c v_s\\
u_s' &= v_s.
\end{aligned}
\right.
\label{eqn:SysTravelWave}
\end{equation}
If $\mathbf{u}^*$ is a Nash equilibrium of $\mathbf{A}$, then the pair $\mathbf{u} = \mathbf{u}^*$ and $\mathbf{v} = \mathbf{0}$ is a fixed point of \cref{eqn:TravelingWave}. Linearizing about $u_r = u_p = u_s = \tfrac{1}{3}$, $v_r = v_p = v_s = 0$, we obtain the eigenvalues:
\begin{align*}
\lambda_{1,2} &= \frac{-3c \pm\sqrt{9c^2+12aD}}{6D}\\
\lambda_{3,4} &= \frac{-3c \pm \sqrt{9c^2 + 6aD + 6D\sqrt{3}\sqrt{-(a+2)^2}}}{6D}\\
\lambda_{5,6} &= \frac{-3c \pm \sqrt{9c^2 + 6aD - 6D\sqrt{3}\sqrt{-(a+2)^2}}}{6D}.\end{align*}
We can simultaneously show that for appropriate choice of wave speed, a Hopf bifurcation and hence a two dimensional center manifold exists and therefore a non-decaying traveling wave solution exists for the PDE. As a by-product, we compute the wave speed for a non-decaying traveling wave in terms of $a$ and $D$. For some constant $b$ (to be determined), let:
\begin{multline*}
(3c \pm bi)^2 = 9c^2 + 6aD \pm 6D\sqrt{3}\sqrt{-(a+2)^2} = \\
9c^2 + i6aD \pm 6D(a+2)\sqrt{3}.
\end{multline*}
Expanding the left hand side and relating real and imaginary parts we have:
\begin{align*}
&9c^2 - b^2 = 9c^2 + 6aD\\
&6bc  = 6D(a+2)\sqrt{3}.
\end{align*} 
Solving for $b$ and $c$ yields:
\begin{align*}
b & = \sqrt{-6aD}\\
\pm\tilde{c} &= \pm\frac{(a+2)\sqrt{2k}}{2\sqrt{-a}}
\end{align*}
Without loss of generality, assume positive $\tilde{c}$. Using this information, we can rewrite the eigenvalues as:
\begin{align*}
\tilde{\lambda}_{1,2} &= \frac{-3\tilde{c} \pm\sqrt{9\tilde{c}^2+12aD}}{6D}\\
\tilde{\lambda}_{3}&= \frac{-6\tilde{c} - bi}{6D}\\
\tilde{\lambda}_{4}&= \frac{bi}{6D}\\
\tilde{\lambda}_{5}&= \frac{-6\tilde{c}+bi}{6D}\\
\tilde{\lambda}_{6}&= \frac{-bi}{6D}.
\end{align*}
For this solution to be physically realized, we must have $a < 0$. We also assume $a > -2$ or the dynamics changes (i.e., winning becomes losing). These requirements and our assumption that $\tilde{c} > 0$ implies that $\mathrm{Re}(\lambda_{1,2}) < 0$ for all choices of $D > 0$ and $a \in (-2,0)$. Therefore, this system has a four dimensional stable manifold and two pure imaginary eigenvalues, which satisfies the first requirement of Hopf's theorem (see \cite{GH13}, Page 152). 

Consider $\lambda_{4,6}$ as functions of $c$ with:
\begin{displaymath}
\lambda_{4,6}(c) = \frac{-3c + \sqrt{9c^2 + 6aD \pm 6D\sqrt{3}\sqrt{-(a+2)^2}}}{6D}.
\end{displaymath}
For $c = \tilde{c} > 0$, we know that $\mathrm{Re}[\lambda_{4,6}(\tilde{c})] = 0$. Differentiating we have:
\begin{displaymath}
\lambda_{4,6}'(c) = - \frac{1}{2D} + \frac{3 c}{2D\sqrt{9c^2 + 6aD \pm 6D\sqrt{3}\sqrt{-(a+2)^2}}}. 
\end{displaymath}
Evaluating at $c = \tilde{c}$ and simpilfying yields:
\begin{displaymath}
\lambda_{4,6}'(\tilde{c}) = - \frac{1}{2D} + \frac{3 \tilde{c}\left(3\tilde{c}\mp bi\right)}{2D\left(9\tilde{c}^2 - b^2\right)},
\end{displaymath}
by choice of $\tilde{c}$. We conclude:
\begin{displaymath}
\mathrm{Re}\left[\lambda_{4,6}'(\tilde{c})\right] = - \frac{1}{2D} + \frac{9\tilde{c}^2}{2D\left(9\tilde{c}^2 - b^2\right)} = \frac{b^2}{2D\left(9\tilde{c}^2 - b^2\right)}.
\end{displaymath}
This is non-zero since we assume $a < 0$ and $D > 0$. Thus the real parts of eigenvalues $\lambda_{4,6}$ cross the imaginary axis with non-zero speed, satisfying the second criterion of Hopf's theorem. Thus we conclude that the six dimensional traveling wave ODE has a solution and moreover exhibits a Hopf bifurcation, implying the existence of traveling wave solutions. In our numeric simulations, we show that fine tuning the parameters leads to a numerically stable traveling wave over the region of integration when $a = -\tfrac{4}{5}$ and $D = \tfrac{1}{12}$ (see \cref{fig:NegBiasPlots,fig:NegativeBiasTernary}).

\subsection{Illustrative Behavior in the Unbiased Games}
\begin{figure}[htbp]
\centering
\includegraphics[width=0.95\columnwidth]{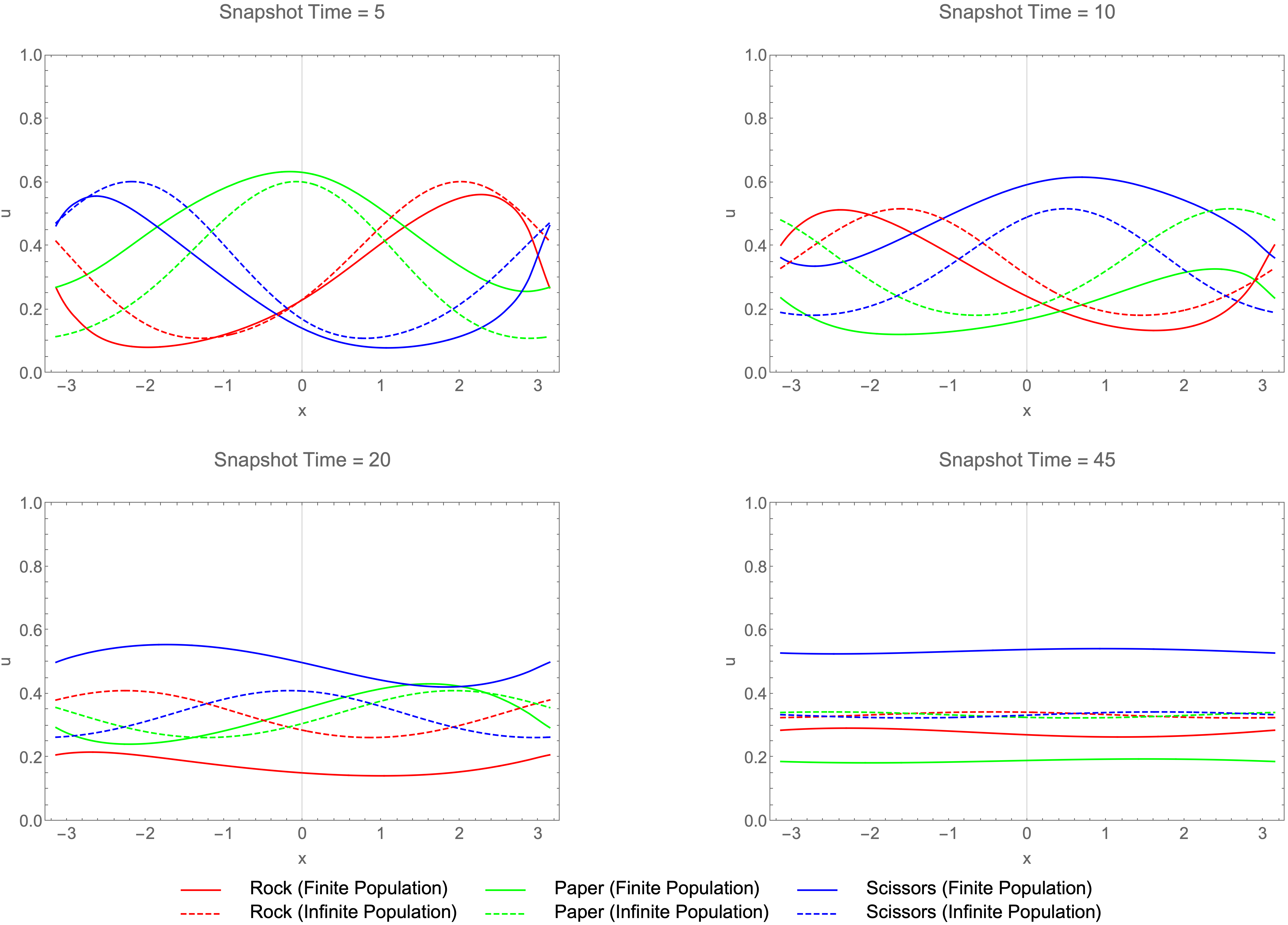}
\caption{Illustration of finite and infinite dynamics at various points in time on the circle for RPS with zero bias. The diffusing population causes finite population to converge to a stationary oscillating solution, while the infinite population converges to a stationary equilibrium solution.}
\label{fig:NoBiasPlots}
\end{figure}

\begin{figure}[htbp]
\centering
\includegraphics[width=0.45\columnwidth]{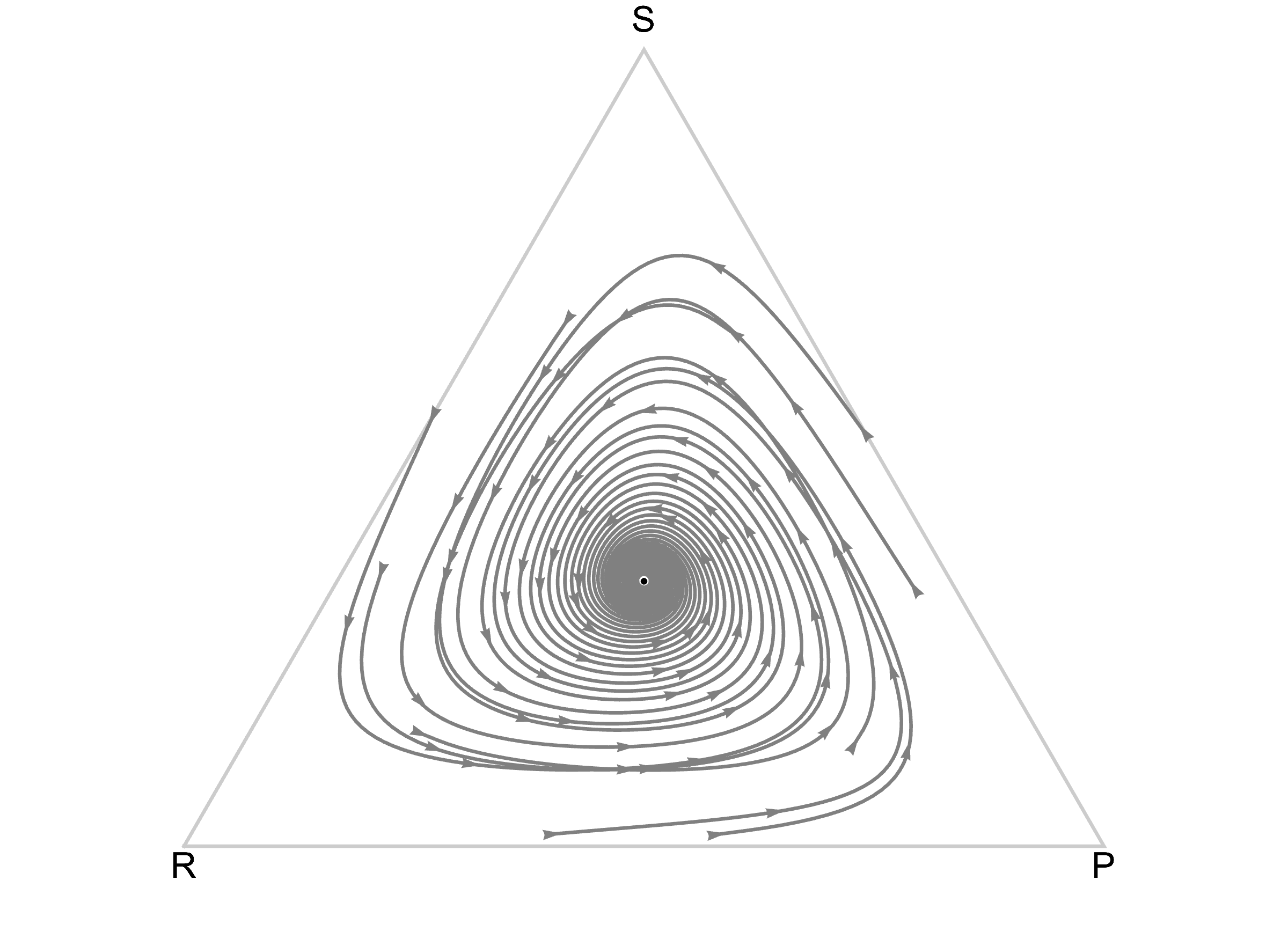} \quad
\includegraphics[width=0.45\columnwidth]{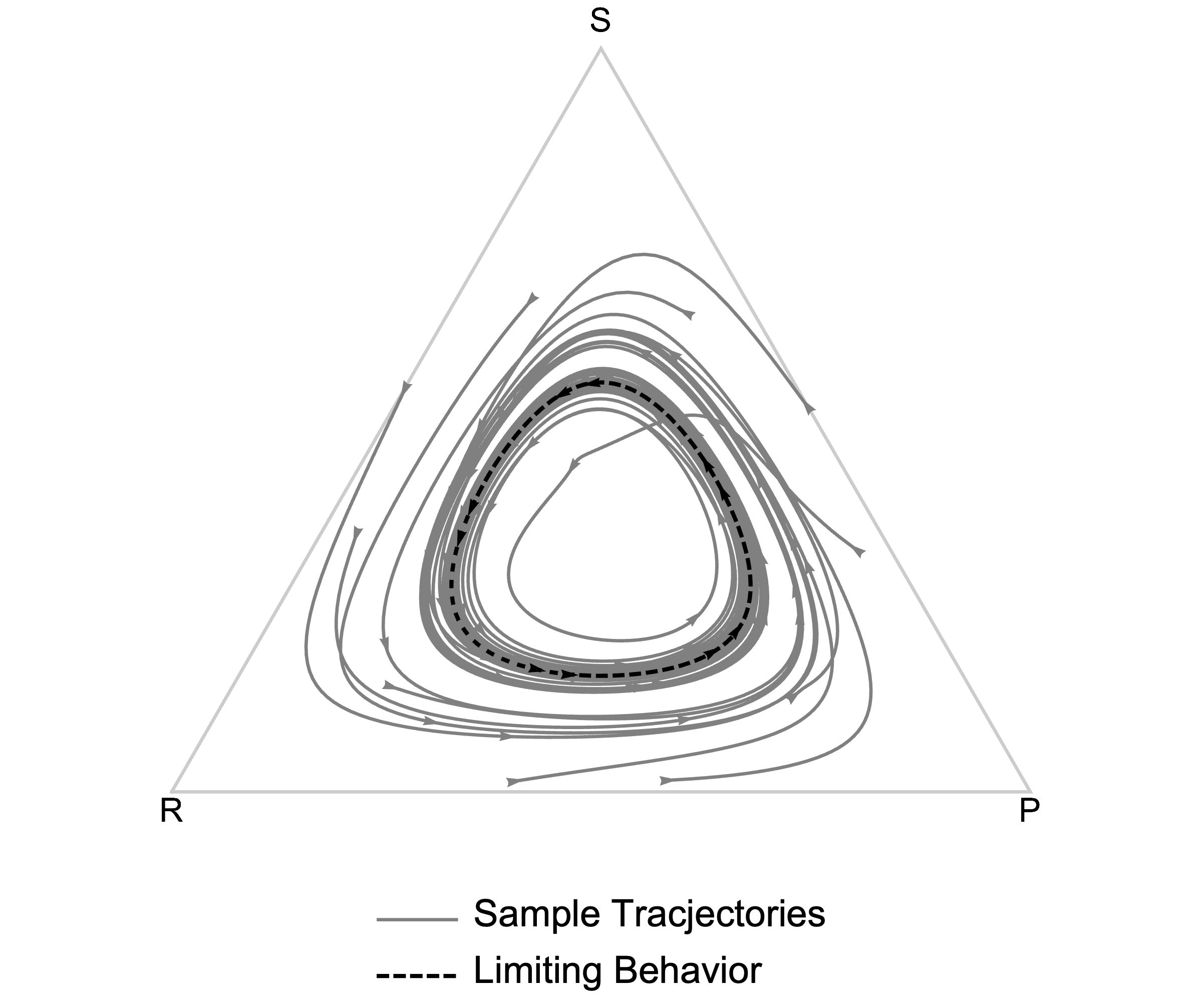}
\caption{(left) The infinite population spatial replicator converges to an equilibrium point as illustrated from multiple starting points on the circle. (right) The finite population spatial replicator converges to an oscillating stationary solution as illustrated from multiple starting points on the circle. The oscillating solution is a solution to the replicator dynamic with unbiased RPS.}
\label{fig:NoBiasTernary}
\end{figure}

Consider \cref{eqn:finitespatialrep,eqn:spatialrep} and assume the periodic boundary conditions $u_{*}(-\pi,t) = u_{*}(\pi,t)$ for all time. When $a = 0$, then $\zeta(u_r,u_p,u_s) = 0$, so the population $M(x,t)$ satisfies the heat equation. For simplicity, we choose a solution to the heat equation that models the diffusion of a population outward:
\begin{displaymath}
M(x,t) = \frac{4000 e^{-\frac{x^2}{4 \beta  (t+10)}}}{\sqrt{\pi } \sqrt{\beta  (t+10)}}.
\end{displaymath}
The initial conditions:
\begin{align*}
u_r(x,0) &= \frac{1}{3}\left[1+\sin\left(x - \tfrac{4\pi}{3}\right)\right]\\
u_p(x,0) &= \frac{1}{3}\left[1+\sin\left(x - \tfrac{2\pi}{3}\right)\right]\\
u_s(x,0) &= \frac{1}{3}\left[1+\sin\left(x\right)\right]
\end{align*}
model three populations that are proportionally spread around a circle. The behavior of the three population proportions in both finite and infinite populations are shown in \cref{fig:NoBiasPlots}.
In the infinite population case we observe that the three population proportions converge toward the equilibrium stationary solution. This is further illustrated in \cref{fig:NoBiasTernary} (left) where we show a ternary plot of the three strategies at $x = 0$. By contrast the finite population solution converges toward the oscillating stationary solution, as illustrated in \cref{fig:NoBiasTernary} (right). We show the corresponding solution to the ordinary RPS replicator dynamic to which the system converges at all points in space. The convergence of the infinite population system to the equilibrium at all points in space is illustrated in \cref{fig:NoBiasTernary} (left). 
Before converging to an oscillating solution, we can see the one-dimensional proportions proportions are affected by the diffusion of the population at large. In \cref{fig:NoBiasPlots}, we note that the finite population plots are stretched with respect to their infinite population counterparts. This is particularly noticeable at time $t = 5$ and $t = 10$. This stretching, caused by the advection of the total population, leads to the difference in steady state solutions for the same initial conditions.

\begin{figure}[htbp]
\centering
\includegraphics[width=0.95\columnwidth]{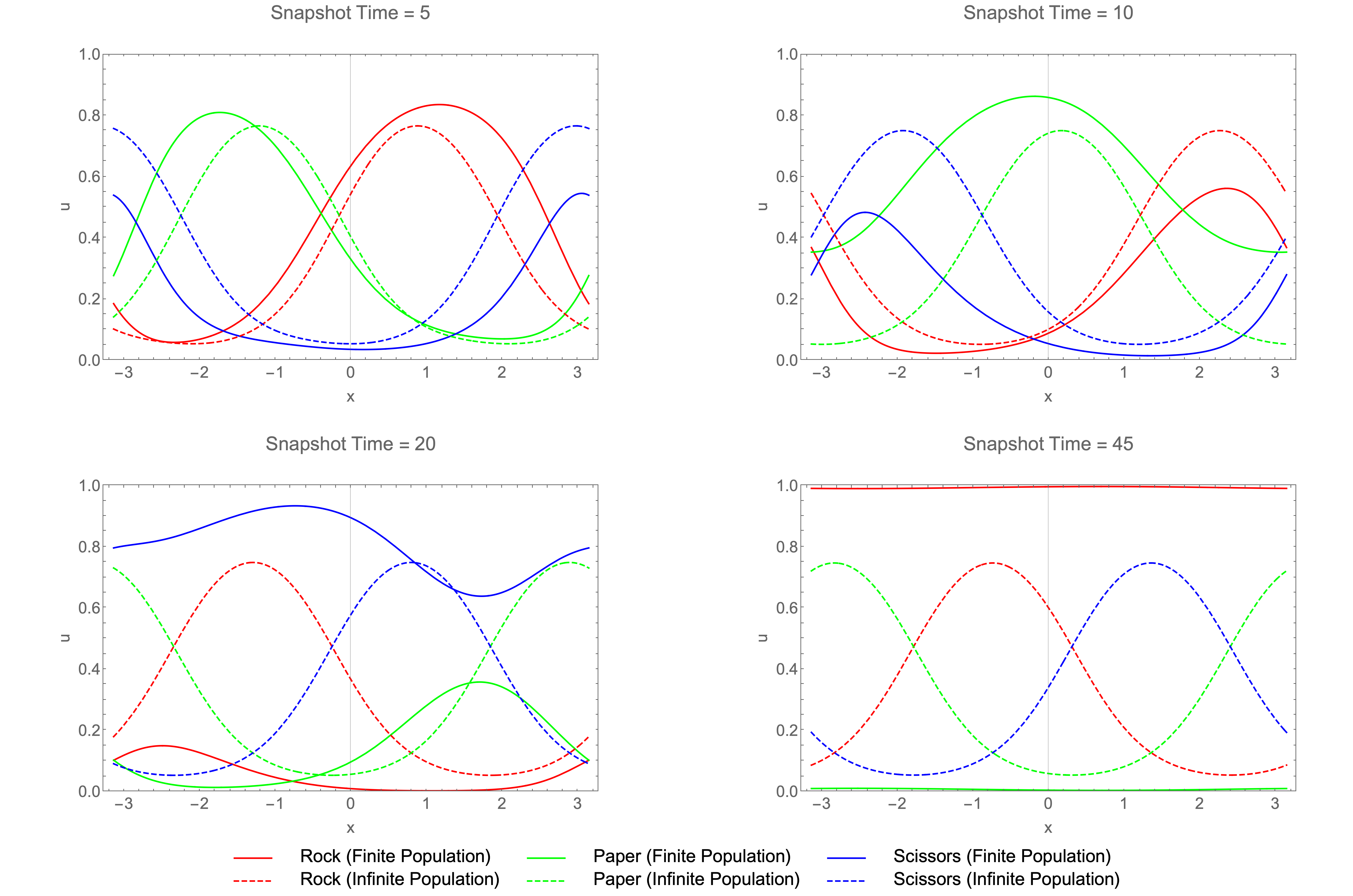}
\caption{Illustration of finite and infinite dynamics at various points in time on the circle for RPS with negative bias. In the infinite population case, a stable traveling wave emerges. In the finite population case, population collapse causes the population proportions to swing with ever increasing amplitude.}
\label{fig:NegBiasPlots}
\end{figure}

\begin{figure}[htbp]
\centering
\includegraphics[width=0.45\columnwidth]{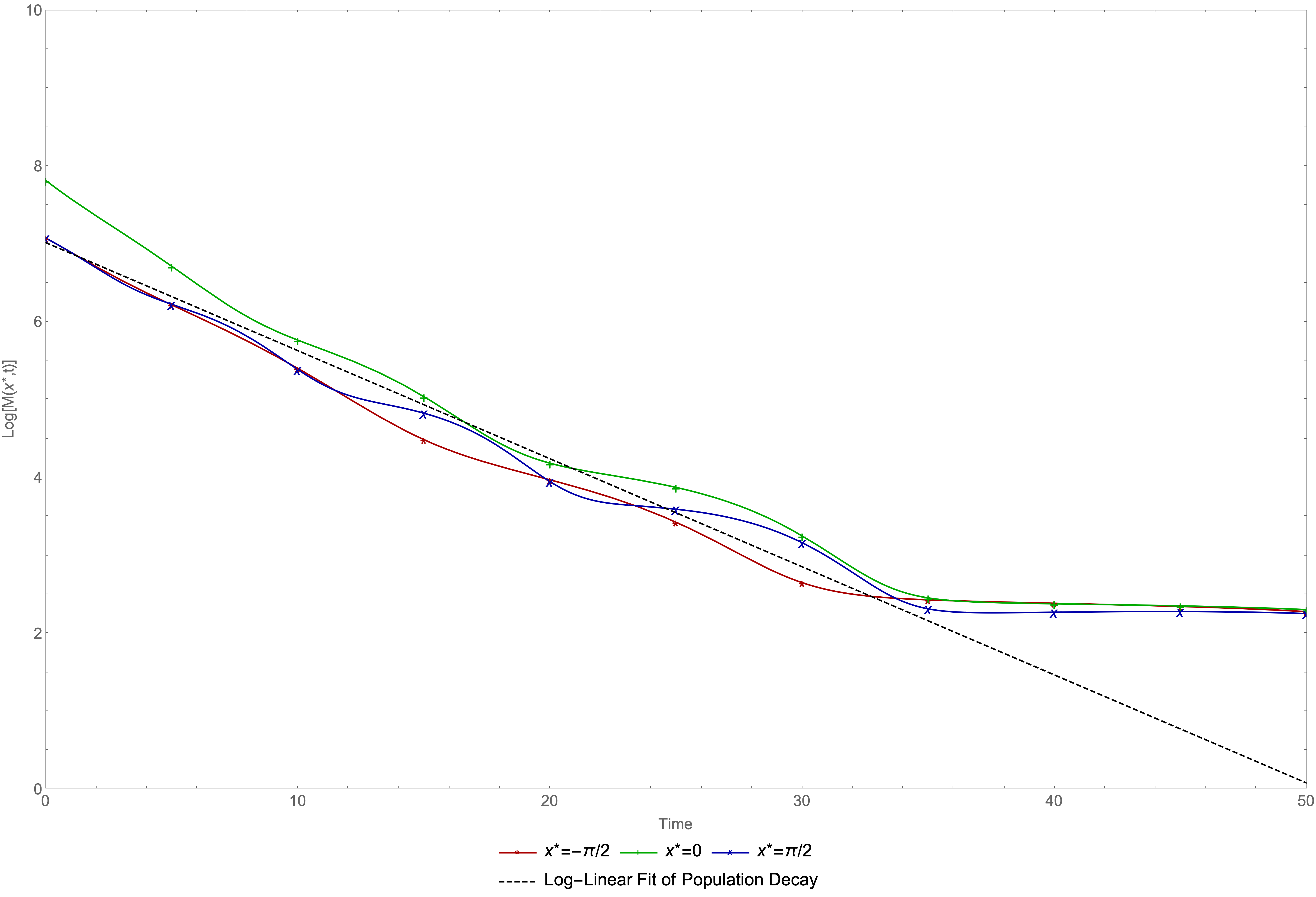}
\caption{The population collapses exponentially until $t \approx 30$. Additionally interactions with the individual species cause the population to become asymmetric as illustrated by the trajectories of $M(x^*,t)$ at $x^* = \pm \tfrac{\pi}{2}$.}
\label{fig:PopulationDecay}
\end{figure}

\begin{figure}[htbp]
\centering
\includegraphics[width=0.45\columnwidth]{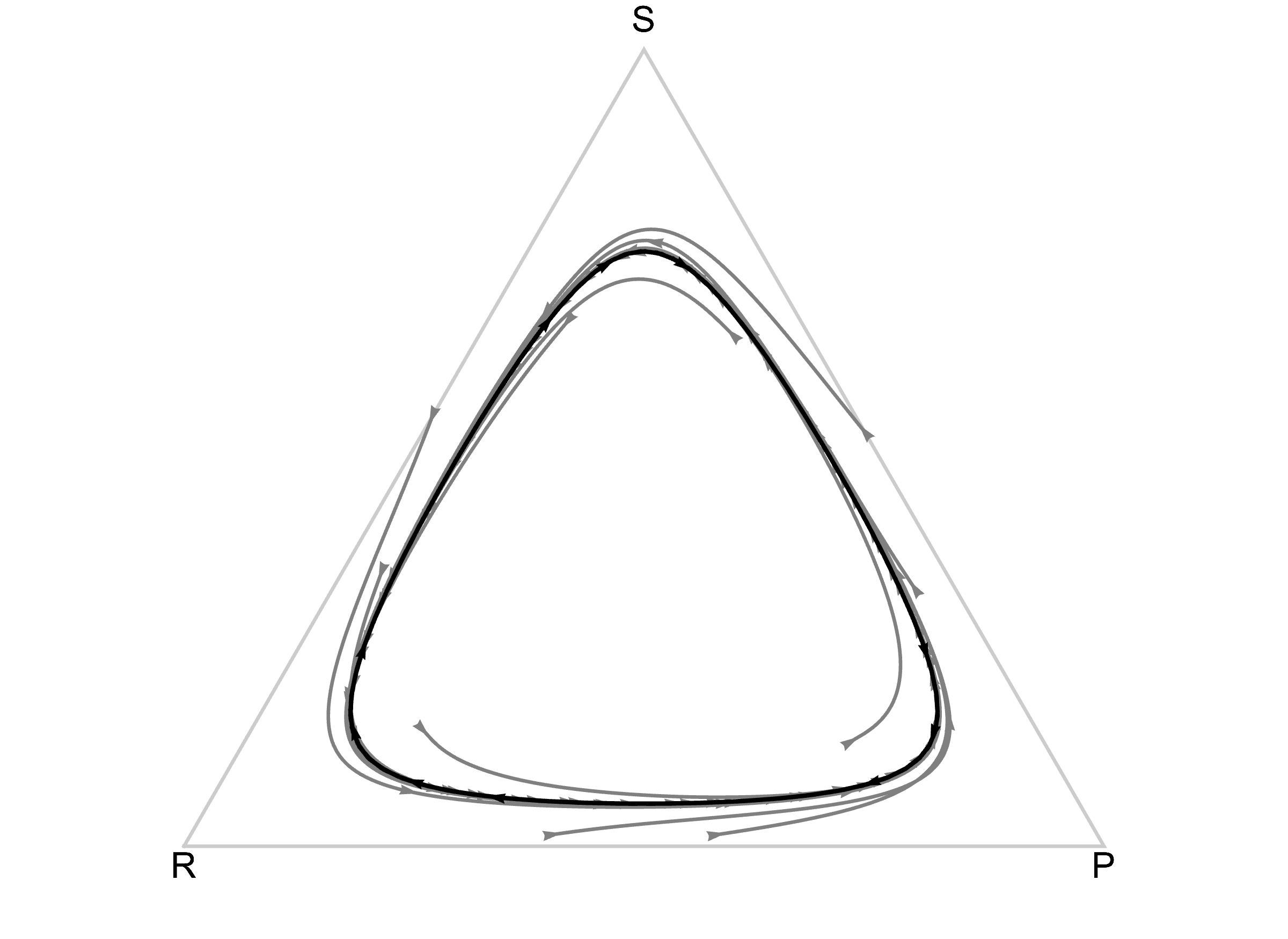}\quad
\includegraphics[width=0.45\columnwidth]{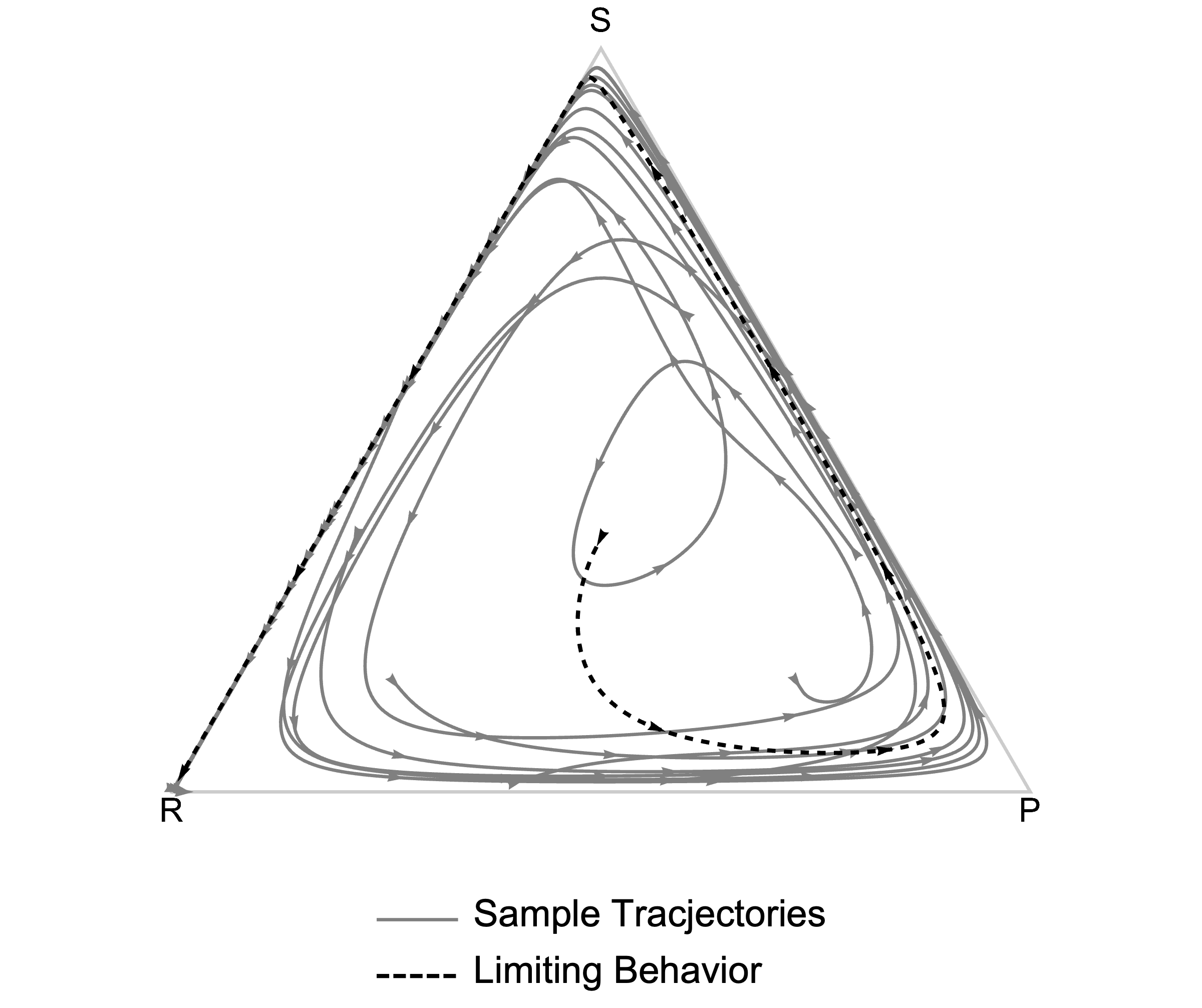}
\caption{(left) The infinite population spatial replicator converges to a limit cycle of the 6 dimensional traveling wave ODE as illustrated from multiple starting points on the circle. (right) The finite population spatial replicator converges to an oscillating stationary solution as illustrated from multiple starting points on the circle. The oscillating solution is a solution to the replicator dynamic with negative biased RPS and thus is converging to the boundary of the simplex.}
\label{fig:NegativeBiasTernary}
\end{figure}

We contrast this behavior with the case when $a < 0$. Here, the population will collapse since $\zeta(u_r,u_p,u_s) \leq 0$ at all times. As noted, we set $a = -\tfrac{4}{5}$ and $D = \tfrac{1}{12}$. From this we compute a constant amplitude traveling wave speed of:
\begin{displaymath}
\tilde{c} = \frac{1}{2}\sqrt{\frac{3}{10}}.
\end{displaymath}
This traveling wave can be seen in \cref{fig:NegBiasPlots} in the infinite population plots. 
In contrast the population collapse (shown in \cref{fig:PopulationDecay}) causes the finite population proportions to converge to an oscillating stationary strategy with greater and greater amplitude. This is precisely the behavior one expects to see from RPS under the replicator dynamics when $a < 0$. This is also shown \cref{fig:NegativeBiasTernary}(right), in which we show an example RPS trajectory with $a = -\tfrac{4}{5}$ and example trajectories at various points in space.
Interestingly, the population collapse slows dramatically as $t$ increases. This is caused by the fact that for (nearly) pure strategies, $\zeta(u_r,u_s,u_t) \approx 0$. This is illustrated in \cref{fig:PopulationDecay}, where the nearly exponential decay flattens  after $t \approx 30$. Additionally, \cref{fig:PopulationDecay} shows that the population becomes asymmetric as the system evolves, as a result of the various species dynamics.

Convergence to the stable traveling wave solution is illustrated in \cref{fig:NegativeBiasTernary} (left). Using the computed $\tilde{c}$, numerical analysis provided initial conditions which can be used to find a numerical solution to  \cref{eqn:SysTravelWave} that produces the closed curve (on the center manifold) that is guaranteed to exist by the eigenvalue analysis performed in the previous section. This is shown in \cref{fig:NegativeBiasTernary} (left). 

\section{Conclusion}
In this paper we studied a finite and infinite population spatial
replicator. We showed how the finite population spatial replicator can
be derived from first principles from a stochastic cellular automaton
model and from there how the infinite population replicator used by
Vickers \cite{ref:vickers1989,ref:vickers1991} follows from this. This result generalizes the work of Durrett and Levin \cite{DL94} who first derived and studied the finite population spatial replicator for a specific game. We then compared the finite and infinite population spatial replicator for rock-paper-scissors on a circle ($S^1$). Our results are consistent with the work in \cite{RMF07,RMF08,HMT10,SMR13,SMR14,SMJS14,PR17,PR19}, which studies various characteristics of RPS, not necessarily in connection with spatial replicator. Consistent with the work in \cite{PR17,PR19} we show that for a certain rock-paper-scissors variant stable amplitude traveling waves can emerge as solutions to the infinite population spatial replicator by proving the existence of a Hopf bifurcation in the traveling wave ODE. These traveling waves are destroyed by population collapse in the finite population spatial replicator.  

The finite population spatial replicator is intriguing because it is a highly non-linear reaction-advection-diffusion equation where advection is governed by the per capita bulk population motion. Identifying cases where complex behaviors emerge in the finite population case is a clear future direction. Additionally, studying stationary solutions may yield insights. For example, in our unbiased RPS, solutions to the stationary population equation are just the harmonic functions. Using this simplification may help identify interesting properties of the population proportion equations. 

\section*{Acknowledgement}
Portions of CG's work were supported by the National Science Foundation Grant CMMI-1932991.

\appendix
\section{Relation to Other Rock Paper Scissors Games}\label{sec:A}
A substantial amount of work has been done on rock-paper-scissors outside of the context of the replicator dynamic. This work does not map conveniently to results on the replicator dynamic or its spatial variants
\cite{PR19,M10,SMR14,SMJS14,RMF07,ML75,RMF08,HMT10,PR17,SMR13,KT21}. The
earliest work to study competition among three species with cyclic
dominance may be \cite{ML75}, which is contemporary but does predate
the earliest work in evolutionary game theory (see e.g.,
\cite{TJ78,Ze80,Hof81,B83}). In the past twenty years, there has been
substantial work on the spatial dynamics of RPS that is independent of
the spatial replicator dynamic
\cite{ref:andrew,ref:kerr,ref:nowak,ref:roca,ref:RPSnetworks,ref:vickers1989,ref:vickers1991,ref:cressmanVickers}.
Work till 2014 is reviewed in \cite{SMJS14}. Mobility in cyclic
competition (RPS) is studied in \cite{RMF07,RMF08}. The impact of
reaction rates on spatial RPS is considered in \cite{HMT10}, while the
emergence of spiraling waves is studied in \cite{SMR13,SMR14}. More
recently, traveling waves, spirals and heteroclininc bifurcations have
been studied in \cite{PR17,PR19}. A discrete time model displaying
chaos is considered in \cite{SAF02}.

The results most closely related to those in this paper (specifically \cref{sec:TravelingWave}) can be found in the pair of papers by Postlethwaite and Rucklidge \cite{PR17,PR19}. Both these papers study a traveling wave solution for a specific spatial model of the rock-paper-scissors, with a more formal treatment given in \cite{PR19}. The motivating aspatial model is given by the system of equations:
\begin{align}
\dot{a} &= a (1-(a+b+c) -(\sigma + \zeta )b + \zeta c) \label{eqn:Leeds1}\\
\dot{b} &= b (1-(a+b+c) -(\sigma + \zeta )c + \zeta a) \label{eqn:Leeds2}\\
\dot{c} &= c (1-(a+b+c) -(\sigma + \zeta )a + \zeta b) \label{eqn:Leeds3}
\end{align}
The observation is made that when $\sigma = 0$, this system exhibits the conserved quantities $a+b+c = 1$ and $abc = K$ for some constant $K$. Letting $u_1 = a$, $u_2 = b$ and $u_3 = c$, then \crefrange{eqn:Leeds1}{eqn:Leeds3} can be written as the standard replicator:
\begin{equation}
\dot{u}_i = u_i\left(\left(\mathbf{e}_i - \mathbf{u}\right)^T\mathbf{A}\mathbf{u} \right),
\label{eqn:LeedsReplicator}
\end{equation}
where:
\begin{equation}
\mathbf{A} = \begin{bmatrix}
-1 & -\zeta -1 & \zeta -1 \\
 \zeta -1 & -1 & -\zeta  -1 \\
 -\zeta -1 & \zeta -1 & -1
\end{bmatrix}
\end{equation}
This is trivially diffeomorphic to the replicator dynamic with payoff matrix:
\begin{equation}
\tilde{\mathbf{A}} = 
\zeta\begin{bmatrix}
0 & -1 & 1 \\
 1 & 0 & -1  \\
 -1 & 1 & 0
\end{bmatrix}.
\end{equation}
This is a scaled version of the standard (unbiased) RPS payoff matrix. However, when $\sigma \neq 0$, then \cref{eqn:Leeds1,eqn:Leeds2,eqn:Leeds3} can be written as:
\begin{equation}
\dot{u}_i = u_i\left(1 + \mathbf{e}_i^T\mathbf{B}\mathbf{u}\right),
\end{equation}
with:
\begin{equation}
\mathbf{B} = \begin{bmatrix}
-1 & -\sigma -\zeta -1 & \zeta -1 \\
 \zeta -1 & -1 & -\sigma -\zeta  -1 \\
 -\sigma -\zeta -1 & \zeta -1 & -1
\end{bmatrix}.
\end{equation}
Population evolution equations of this form are considered in \cite{PG16}.
When $\sigma = 0$, then $\mathbf{u}^T\mathbf{B}\mathbf{u} = \mathbf{u}^T\mathbf{A}\mathbf{u} = -1$, from which we derive \cref{eqn:LeedsReplicator}. For $\sigma \neq 0$, this is not the case and consequently the dynamics of \cref{eqn:Leeds1,eqn:Leeds2,eqn:Leeds3} are not trapped on the unit 2-simplex as will be the case when we study (un)biased spatial replicator dynamics in finite and infinite populations. To be clear, the authors of \cite{PR17,PR19} make no claim to this effect, however it does create an important distinction between the work in these papers and the work in this paper.
\begin{figure}[htbp]
\centering
\includegraphics[width=0.45\columnwidth]{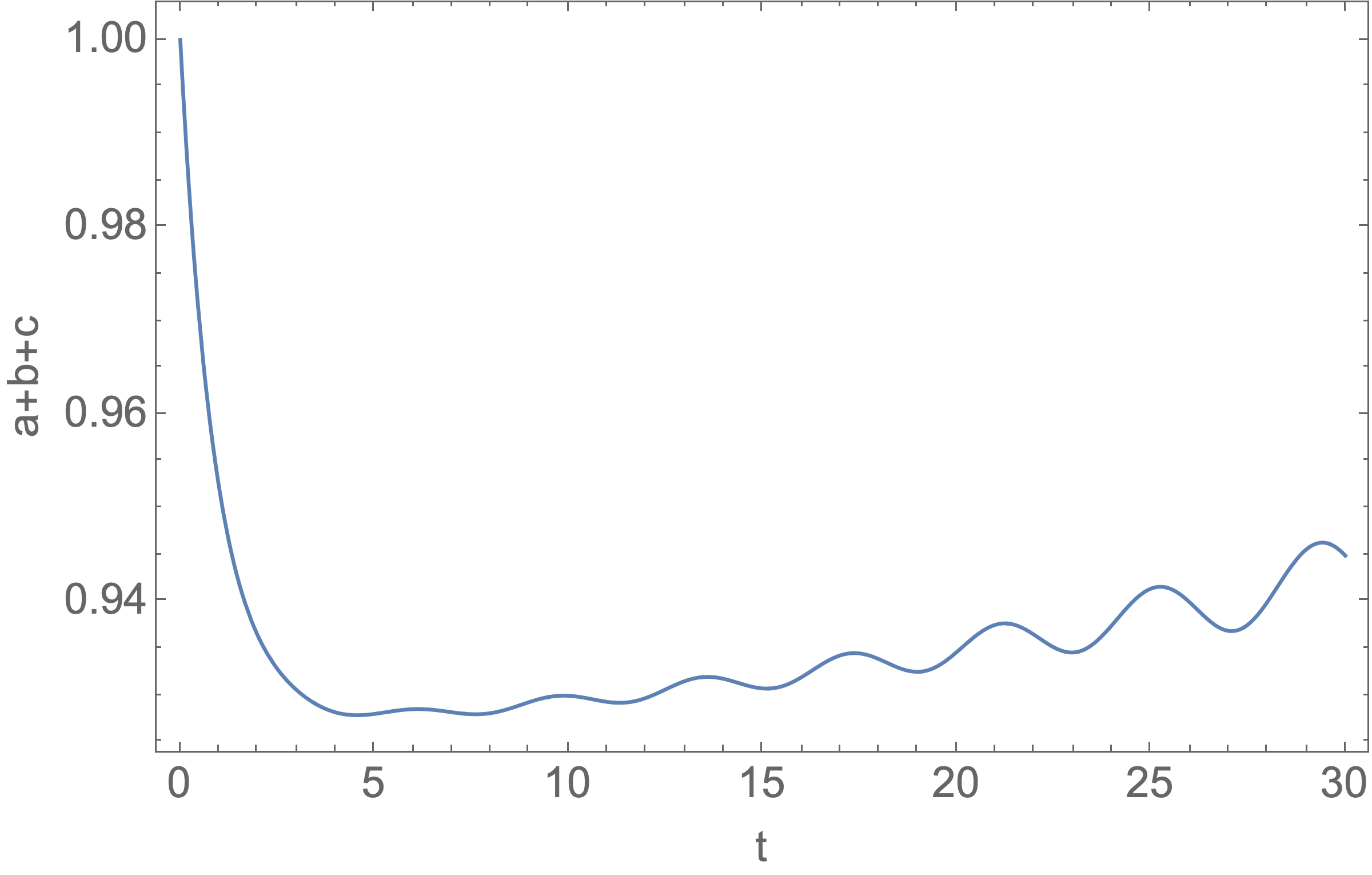}
\caption{We illustrate that the dynamics studied in \cite{PR17,PR19} are not trapped on the unit simplex for $\sigma \neq 0$, while the same dynamics are illustrative of an unbiased RPS game when $\sigma = 0$.}
\label{fig:Leeds}
\end{figure}
Additionally, as illustrated in \cref{fig:Leeds}, the ODE dynamics that give rise to the PDE are not trapped on the unit simplex (as they are in the replicator) thus suggesting distinct behaviors may be observed.

\bibliographystyle{unsrt}
\bibliography{mrabbrev,spatialreplicator,RevisionBib}

\end{document}